\documentclass[aps,pra,preprint,groupedaddress]{revtex4}

\begin{document}


\title{Pair density functional theory by means of the correlated wave 
function}


%
\author{Masahiko Higuchi}
\affiliation{Department of Physics, Faculty of Science, 
Shinshu University, Matsumoto 390-8621, Japan}
\author{Katsuhiko Higuchi}
\affiliation{Graduate School of Advanced Sciences of Matter, 
Hiroshima University, Higashi-Hiroshima 739-8527, Japan}
%

\date{\today}

\begin{abstract}
We present a density functional scheme for calculating the pair density (PD) 
by means of the correlated wave function. This scheme is free from both of 
problems related to PD functional theory, i.e., (a) the need to constrain 
the variational principle to $N$-representable PDs and 
(b) the development of a kinetic energy functional. 
By using the correlated wave function, the searching region 
for the ground-state PD is substantially extended as compared with 
our previous theory[Physica B \textbf{372} (2007), in press]. 
The variational principle results in the simultaneous equations that yield 
the best PD beyond the previous theory, not to mention the Hartree-Fock 
approximation. 
 \\
 \\
 \\
Note: The previous paper [Physica B \textbf{372 }(2007), in press] can be 
downloaded from ``Article in press'' in the website of Physica B. 
\end{abstract}

\pacs{71.15.Mb , 31.15.Ew, 31.25.Eb}
\maketitle
%
%
%
\section{INTRODUCTION}
The pair density (PD) functional theory has recently attracted particular 
interest because it provides the obvious way to improve on the density 
functional theory (DFT)\cite{1,2,3,4,5,6,7,8,9}.  Ziesche first proposed 
the PD functional theory about a decade ago\cite{1,2}, 
and then many workers followed his work and 
have developed a variety of approaches\cite{3,4,5,6,7,8,9}.

Very recently, we have proposed an approximate scheme for calculating the PD 
on the basis of the extended constrained-search theory\cite{10,11,12,13,14}. 
By introducing the noninteracting reference system\cite{10,11}, 
the resultant PD corresponds to the best solution within the set of the PDs 
that are constructed from the single Slater determinant (SSD). 
This PD functional theory has two kinds of merits. The first one is 
that the reproduced PD is necessarily $N$-representable. This is a strong 
merit because the necessary and sufficient conditions for 
the $N$-representable PD have not yet been 
known\cite{15,16,17,18,19,20,21,22,23,24,25,26}. 
The second merit is the tractable form of the kinetic energy 
functional. The kinetic energy functional cannot exactly be written by using 
the PD alone. Some approximation is required to be introduced\cite{7}. In this 
theory, we have successfully given an approximate form of the kinetic energy 
functional with the aid of the coordinate scaling of electrons\cite{10,11}.

On the other hand, we also have the remaining problem in it\cite{10}. Namely, 
there exists the possibility that the solution might be far from the correct 
value of the ground-state PD. This is because the searching region of the 
PDs may be smaller than the set of $N$-representable PDs. In order to improve 
the PD functional theory with keeping the above-mentioned merits, we have to 
extend the searching region of the PDs to the set of $N$-representable PDs as 
closely as possible. At least, we had better extend the searching region 
beyond the set of the SSD-representable PDs\cite{27}.

In this paper, we shall employ the strategy for reproducing the PDs not by 
means of the SSD, but through the correlated wave function. As the 
correlated wave function, we adopt the Jastrow wave function that is defined 
as the SSD multiplied by the correlation function\cite{28,29}.  Owing to the 
correlation function, the searching region substantially becomes larger than 
the set of the SSD-representable PDs. Of course, the reproduced PD is kept 
$N$-representable because the PD is calculated via the Jastrow wave function 
that is a kind of antisymmetric wave functions. Also the second merit is not 
missed in the present scheme by utilizing the result of the scaling property 
of the kinetic energy functional. 

Organization of this paper is as follows. In Sec. II, we provide the 
preliminary definitions of various quantities that appear in the following 
sections. In Sec. III, by means of the variational principle with respect to 
the PD, we derive simultaneous equations that yield the best PD that is 
superior to the previous one\cite{10}. Such equations are quite tractable, 
the computational method of which are also proposed in Sec. III.

\section{PRELIMINARY DEFINITIONS IN THE PD FUNCTIONAL THEORY}
In this section we give the preliminary definitions which will be used in 
the present scheme. The PD is defined as the diagonal elements of the 
spinless second-order reduced density matrix, i.e., 
\begin{equation}
\label{eq1}
\gamma ^{(2)} ({\rm {\bf r{r}'}};{\rm {\bf r{r}'}})=\left\langle \Psi 
\right|\frac{1}{2}\int\!\!\!\int {\hat {\psi }^{+}({\rm {\bf r}},\,\eta 
)\hat {\psi }^{+}({\rm {\bf {r}'}},\,{\eta }')\hat {\psi }({\rm {\bf 
{r}'}},\,{\eta }')\hat {\psi }({\rm {\bf r}},\,\eta )} \mbox{d}\eta 
\mbox{d}{\eta }'\left| \Psi \right\rangle ,
\end{equation}
where $\hat {\psi }({\rm {\bf r}},\,\eta )$ 
and $\hat {\psi }^{+}({\rm {\bf r}},\,\eta )$ 
are field operators of electrons, and $\Psi $ is the 
antisymmetric wave function, and ${\rm {\bf r}}$ and $\eta $ are spatial and 
spin coordinates, respectively. We shall consider a system, the Hamiltonian 
of which is given by
\begin{equation}
\label{eq2}
\hat {H}=\hat {T}+\hat {W}+\int {\mbox{d}{\rm {\bf r}}\hat {\rho }({\rm {\bf 
r}})v_{ext} ({\rm {\bf r}})} ,
\end{equation}
where $\hat {T}$, $\hat {W}$ and $\hat {\rho }({\rm {\bf r}})$ are operators 
of the kinetic energy, electron-electron interaction and electron density, 
respectively, and $v_{ext} ({\rm {\bf r}})$ stands for the external 
potential. In the similar way to the extended constrained-search 
theory\cite{12,13,14}, the universal functional is defined as
\begin{eqnarray}
\label{eq3}
F[\gamma ^{(2)}]&=&
\mathop {\mbox{Min}}\limits_{\Psi \to \gamma^{(2)}
({\rm {\bf r{r}'}};{\rm {\bf r{r}'}})} \left\langle \Psi 
\right|\hat {T}+\hat {W}\left| \Psi \right\rangle \nonumber \\ 
&=&\left\langle {\Psi [\gamma^{(2)}]} \right|\hat {T}+\hat {W}\left| 
{\Psi [\gamma ^{(2)}]} \right\rangle, 
\end{eqnarray}
where $\Psi \to \gamma ^{(2)} ({\rm {\bf r{r}'}};{\rm {\bf r{r}'}})$ 
denotes the searching over all antisymmetric wave functions that yield a 
prescribed $\gamma ^{(2)} ({\rm {\bf r{r}'}};{\rm {\bf r{r}'}})$. In the 
second line, the minimizing wave function is expressed as $\Psi [\gamma 
^{(2)}]$. By using Eq. (\ref{eq3}), the Hohenberg-Kohn theorems for the PD 
functional theory can be easily proved\cite{1,5}.  Here we show only their 
results\cite{10}: 
\begin{equation}
\label{eq4}
\Psi _{0} =\Psi [\gamma _{0}^{(2)} ],
\end{equation}
and
\begin{eqnarray}
\label{eq5}
 E_{0} &=&\mathop {\mbox{Min}}\limits_{\gamma ^{(2)}} E[\gamma ^{(2)}] 
          \nonumber \\ 
       &=& E[\gamma _{0}^{(2)} ], 
\end{eqnarray}
where $\Psi _{0}$, $E_{0}$ and $\gamma _{0}^{(2)}$ are the 
ground-state wave function, ground-state energy and ground-state PD, 
respectively, and where $E[\gamma ^{(2)}]$ is the total energy functional 
that is given by
\begin{equation}
\label{eq6}
E[\gamma ^{(2)}]=F[\gamma ^{(2)}]+\frac{2}{N-1}\int\!\!\!\int {\mbox{d}{\rm 
{\bf r}}\mbox{d}{\rm {\bf {r}'}}v_{ext} ({\rm {\bf r}})\gamma 
^{(2)}({\rm {\bf r{r}'}};{\rm {\bf r{r}'}})} .
\end{equation}
Equations (\ref{eq4}) and (\ref{eq5}) correspond to the first 
and second Hohenberg-Kohn theorems, respectively. Let us suppose that
\begin{equation}
\label{eq7}
T[\gamma ^{(2)}]=\left\langle {\Psi [\gamma ^{(2)}]} \right|\hat {T}\left| 
{\Psi [\gamma ^{(2)}]} \right\rangle ,
\end{equation}
then Eq. (\ref{eq6}) is rewritten as
\begin{equation}
\label{eq8}
E[\gamma ^{(2)}]=T[\gamma ^{(2)}]+e^{2}\int\!\!\!\int {\mbox{d}{\rm {\bf 
r}}\mbox{d}{\rm {\bf {r}'}}\frac{\gamma ^{(2)}({\rm {\bf r{r}'}};{\rm {\bf 
r{r}'}})}{\left| {{\rm {\bf r}}-{\rm {\bf {r}'}}} \right|}} 
+\frac{2}{N-1}\int\!\!\!\int {\mbox{d}{\rm {\bf r}}\mbox{d}{\rm {\bf 
{r}'}}v_{ext} ({\rm {\bf r}})\gamma ^{(2)}({\rm {\bf r{r}'}};{\rm 
{\bf r{r}'}}),} 
\end{equation}
where, in the second term, we use the fact that the expectation value of 
$\hat {W}$ is exactly written in terms of 
$\gamma ^{(2)} ({\rm {\bf r{r}'}};{\rm {\bf r{r}'}})$. 
Equation (\ref{eq8}) is the starting expression for the 
total energy functional in the PD functional theory.

As mentioned in Sec. I, the kinetic energy of the PD functional theory 
cannot be exactly expressed by the PDs alone. In other words, we have to 
employ the approximate form in Eq. (\ref{eq8}). So far, the kinetic energy 
functional of the PD functional theory has been developed by several 
workers\cite{10,30,31}.  In this paper, we make use of an approximate form of 
the kinetic energy functional which has been derived by utilizing the 
scaling property of the kinetic energy functional\cite{10,31}. 
The explicit form is given by
\begin{equation}
\label{eq9}
T[\gamma ^{(2)}]=K\int\!\!\!\int {\mbox{d}{\rm {\bf r}}\mbox{d}{\rm {\bf 
{r}'}}\gamma ^{(2)}({\rm {\bf r{r}'}};{\rm {\bf r{r}'}})^{\frac{4}{3}}} ,
\end{equation}
where $K$ is an arbitrary constant. 

\section {SINGLE-PARTICLE EQUATIONS}
Equation (\ref{eq5}) corresponds to the variational principle 
with respect to the PD. The searching region of the PDs should be of 
course within the set of $N$-representable PDs. 
For that purpose, we shall introduce the searching 
region of the PDs that are calculated from the correlated wave functions. 
The searching region is substantially extended as compared with the previous 
theory\cite{10}, because it is restricted within the set of SSD-representable 
PDs. Extension of the searching region can be regarded as one of appropriate 
developments of the PD functional theory\cite{27}. 

In this paper we adopt the Jastrow wave function as the correlated wave 
function. The explicit evaluation of the PD using the Jastrow wave function 
is actually very hard\cite{28,29}.  As a consequence, the approximation 
technique to evaluate the PD has been developed especially 
in the field of nuclear physics. The expectation value of the PD operator 
with respect to the Jastrow wave function can be systematically expressed 
with the aid of the Yvon-Mayer diagrammatic technique\cite{28,29}.  
Here we shall use the lowest-order approximation of the expectation value 
of the PD operator.

The Jastrow wave function is defined as \cite{28,29}
\begin{equation}
\label{eq10}
\Psi _{J} (x_{1} ,x_{2} ,\cdot \cdot \cdot \cdot \cdot ,x_{N} 
)=\frac{1}{\sqrt {C_{N} } }\prod\limits_{1\le i<j\le N} {f(r_{ij} )\Phi 
_{SSD} (x_{1} ,x_{2} ,\cdot \cdot \cdot \cdot \cdot ,x_{N} ),} 
\end{equation}
where $\Phi _{SSD} (x_{1} ,x_{2} ,\cdot \cdot \cdot \cdot \cdot ,x_{N} )$ is 
the SSD, and where 
$f(r_{ij} )=
f\left( {\left| {{\rm {\bf r}}_{i} -{\rm {\bf r}}_{j} } \right|} \right)$ 
is the correlation function, 
and where $C_{N} $ is the normalization constant. 
Suppose that the correlation function is 
chosen to satisfy the cusp condition for the antisymmetric wave function. 
The lowest-order approximation for the expectation value of the PD operator 
is given by \cite{28}
\begin{equation}
\label{eq11}
\gamma ^{(2)} ({\rm {\bf r{r}'}};{\rm {\bf r{r}'}})=\left| {f\left( 
{\left| {{\rm {\bf r}}-{\rm {\bf {r}'}}} \right|} \right)} \right|^{2}\gamma 
_{SSD}^{(2)} ({\rm {\bf r{r}'}};{\rm {\bf r{r}'}}),
\end{equation}
where $\gamma _{SSD}^{(2)} ({\rm {\bf r{r}'}};{\rm {\bf r{r}'}})$ is the 
expectation value of the PD operator with respect to the SSD. Supposing $N$ 
orthonormal spin orbitals of the SSD are denoted as $\left\{ {\psi _{\mu } 
(x)} \right\}$, then Eq. (\ref{eq11}) is explicitly expressed as
\begin{eqnarray}
\label{eq12}
\gamma ^{(2)} ({\rm {\bf r{r}'}};{\rm {\bf r{r}'}})=\frac{1}{2}\left| 
{f\left( {\left| {{\rm {\bf r}}-{\rm {\bf {r}'}}} \right|} \right)} 
\right|^{2}\sum\limits_{\mu _{1} ,\mu _{2} =1}^{N} {\int\!\!\!\int 
{\mbox{d}\eta \mbox{d}{\eta }'
\left\{ {\psi _{\mu _{1} }^{\ast } (x)\psi _{\mu _{2} }^{\ast } 
({x}')\psi _{\mu _{1} } (x)\psi _{\mu _{2} } ({x}')} \right.} } & &
\nonumber \\ 
-\left. {\psi _{\mu _{1} }^{\ast } (x)\psi _{\mu _{2} }^{\ast } ({x}')
\psi _{\mu _{2} } (x)\psi _{\mu _{1} } ({x}')} \right\}&.&
\end{eqnarray}
Next, let us consider the variational principle with respect to the PD, i.e. 
Eq. (\ref{eq5}). The variation of the PD is performed via the spin 
orbitals of Eq. (\ref{eq12}) with the restriction that they 
are orthonormal to each other. Using the Lagrange method of undetermined 
multipliers, we minimize the following functional without the restriction:
\begin{equation}
\label{eq13}
\Omega \left[ {\left\{ {\psi _{\mu } } \right\}} \right]=E\left[ {\gamma 
^{(2)} } \right]-\sum\limits_{\mu ,\nu } {\varepsilon _{\mu \nu } \left\{ 
{\int {\psi _{\mu }^{\ast } (x)\psi _{\nu } (x)\mbox{d}x} 
-\delta _{\mu \nu } } 
\right\}} ,
\end{equation}
where Eqs. (\ref{eq8}), (\ref{eq9}) and (\ref{eq12}) are used 
in the first term on the right-hand side. The minimizing condition 
$\delta \Omega \left[ {\left\{ {\psi _{\mu } } \right\}} \right]=0$ 
immediately leads to
\begin{eqnarray}
\label{eq14}
 & &\sum\limits_\nu {\int {\mbox{d}x_{1} \left\{ {\psi _{\nu }^{\ast }(x_{1} )
 \psi_{\nu } (x_{1} )\psi _{\mu } (x)-\psi _{\nu }^{\ast } (x_{1} )\psi 
_{\nu } (x)\psi _{\mu } (x_{1} )} \right\}} } \nonumber \\ 
 &\times& \left| {f\left( {\left| {{\rm {\bf r}}-{\rm {\bf r}}_{1} } \right|} 
\right)} \right|^{2}\left\{ {\frac{4K}{3}\gamma ^{(2)}({\rm {\bf rr}}_{1} 
;{\rm {\bf rr}}_{1} )^{\frac{1}{3}}+\frac{e^{2}}{\left| {{\rm {\bf r}}-{\rm 
{\bf r}}_{1} } \right|}+\frac{1}{N-1}\left( {v_{ext} ({\rm {\bf r}})+v_{ext} 
({\rm {\bf r}}_{1} )} \right)} \right\} \nonumber \\ 
 &=&\sum\limits_\nu {\varepsilon _{\mu \nu } \psi _{\nu } (x)} , 
\end{eqnarray}
where the chain rule for the functional derivatives is utilized. The 
Lagrange multipliers $\varepsilon _{\mu \nu } $ should be determined by 
requiring that the spin orbitals are orthonormal to each other:
\begin{equation}
\label{eq15}
\int {\psi _{\mu }^{\ast } (x)\psi _{\nu } (x)\mbox{d}x} =\delta _{\mu \nu }.
\end{equation}
In the similar way to the previous theory\cite{10}, we can simplify the above 
equations by means of a unitary transformation of the spin orbitals. It is 
easily shown that $\varepsilon _{\mu \nu } $ forms the Hermitian matrix. 
Suppose that the unitary matrix which diagonalizes $\varepsilon _{\mu \nu } 
$ is written by $U_{\mu \nu } $, then
\begin{equation}
\label{eq16}
\sum\limits_{i,j} {U_{i\mu } ^{\ast }\varepsilon _{ij} U_{j\nu } } =\tilde 
{\varepsilon }_{\mu } \delta _{\mu \nu } 
\end{equation}
is satisfied, where $\tilde {\varepsilon }_{\mu } $ is the diagonal element 
of the diagonal matrix. Let us consider the 
following transformation of the spin orbitals:
\begin{equation}
\label{eq17}
\psi _{\mu } (x)=\sum\limits_\nu {U_{\mu \nu } \chi _{\nu } (x)} .
\end{equation}
Substituting Eq. (\ref{eq17}) into Eq. (\ref{eq14}), 
and using Eq. (\ref{eq16}), we obtain
\begin{eqnarray}
\label{eq18}
& &\sum\limits_\nu {\int {\mbox{d}x_{1} \left\{ {\chi _{\nu }^{\ast } (x_{1} )
\chi_{\nu } (x_{1} )\chi _{\mu } (x)-\chi _{\nu }^{\ast } (x_{1} )\chi 
_{\nu } (x)\chi _{\mu } (x_{1} )} \right\}} } \nonumber \\ 
&\times& \left| {f\left( {\left| {{\rm {\bf r}}-{\rm {\bf r}}_{1} } \right|} 
\right)} \right|^{2}\left\{ {\frac{4K}{3}\gamma ^{(2)}({\rm {\bf rr}}_{1} 
;{\rm {\bf rr}}_{1} )^{\frac{1}{3}}+\frac{e^{2}}{\left| {{\rm {\bf r}}-{\rm 
{\bf r}}_{1} } \right|}+\frac{1}{N-1}\left( {v_{ext} ({\rm {\bf r}})+v_{ext} 
({\rm {\bf r}}_{1} )} \right)} \right\} \nonumber \\ 
&=& \tilde {\varepsilon }_{\mu } \chi _{\mu } (x). 
\end{eqnarray}
Also, Eq. (\ref{eq15}) is transformed into 
\begin{equation}
\label{eq19}
\int {\chi _{\mu }^{\ast } (x)\chi _{\nu } (x)\mbox{d}x} =\delta _{\mu \nu }.
\end{equation}
Here note that the expression for 
$\gamma ^{(2)} ({\rm {\bf r{r}'}};{\rm {\bf r{r}'}})$ 
in Eq. (\ref{eq18}) is kept invariant under the unitary 
transformation. This is confirmed by substituting Eq. (\ref{eq17}) 
into Eq. (\ref{eq12}), i.e.,
\begin{eqnarray}
\label{eq20}
 \gamma ^{(2)} ({\rm {\bf r{r}'}};{\rm {\bf r{r}'}})=\frac{1}{2}\left| 
{f\left( {\left| {{\rm {\bf r}}-{\rm {\bf {r}'}}} \right|} \right)} 
\right|^{2}\sum\limits_{\mu _{1} ,\mu _{2} =1}^{N} {\int\!\!\!\int 
{\mbox{d}\eta \mbox{d}{\eta }'
\left\{ {\chi _{\mu _{1} }^{\ast } (x)\chi _{\mu _{2} }^{\ast } 
({x}')\chi _{\mu _{1} } (x)\chi _{\mu _{2} } ({x}')} \right.} }& & 
\nonumber \\ 
-\left. {\chi _{\mu _{1} }^{\ast } (x)\chi _{\mu _{2} }^{\ast } ({x}')
\chi _{\mu _{2} } (x)\chi _{\mu _{1} } ({x}')} \right\}&.& 
\end{eqnarray}
Equations (\ref{eq18}) and (\ref{eq19}) are the simultaneous equations, 
and the solutions yield the best PD within the set of PDs that 
are calculated from Eq. (\ref{eq20}).

Our previous work may be the first proposal of a computational approach that 
deals with problems related to the PD functional theory\cite{10}. The present 
scheme is also a computational approach, and further improves on the 
previous theory concerning the searching region of the PDs. In that sense, 
it would be useful to consider a computational procedure for solving the 
simultaneous equations (\ref{eq18}) and (\ref{eq19}).

The procedure proposed here is similar to that of the Hartree-Fock 
equation\cite{32}. In order to make the computational procedure readily 
comprehensible, let us rewrite Eq. (\ref{eq18}) as
\begin{equation}
\label{eq21}
\left\{ {F({\rm {\bf r}})-\tilde {\varepsilon }_{\delta } } \right\}\chi 
_{\delta } (x)=G_{\delta } (x)
\end{equation}
with
\begin{eqnarray}
\label{eq22}
 F({\rm {\bf r}})=\int \!\!&{\mbox{d}x_{1}}&\!\! 
 \left| {f\left( {\left| {{\rm {\bf r}}-{\rm 
{\bf r}}_{1} } \right|} \right)} \right|^{2}\sum\limits_{\nu =1}^{N} {\left| 
{\chi _{\nu } (x_{1} )} \right|^{2}} \nonumber \\ 
\!\!&\times&\!\! \left\{ {\frac{4K}{3}\gamma ^{(2)}
({\rm {\bf rr}}_{1} ;{\rm {\bf 
rr}}_{1} )^{\frac{1}{3}}+\frac{e^{2}}{\left| {{\rm {\bf r}}-{\rm {\bf 
r}}_{1} } \right|}+\frac{1}{N-1}\left( {v_{ext} ({\rm {\bf r}})+v_{ext} 
({\rm {\bf r}}_{1} )} \right)} \right\}, 
\end{eqnarray}
\begin{eqnarray}
\label{eq23}
 G_{\delta } (x)=\int \!\!&{\mbox{d}x_{1}}&\!\! 
 \left| {f\left( {\left| {{\rm {\bf r}}-{\rm 
{\bf {r}'}}} \right|} \right)} \right|^{2}\left\{ {\sum\limits_{\nu =1}^{N} 
{\chi _{\nu }^{\ast } (x_{1} )\chi _{\delta } (x_{1} )\chi _{\nu } 
(x)} } \right\} \nonumber \\ 
\!\!&\times&\!\! \left\{ {\frac{4K}{3}\gamma ^{(2)}
({\rm {\bf rr}}_{1} ;{\rm {\bf 
rr}}_{1} )^{\frac{1}{3}}+\frac{e^{2}}{\left| {{\rm {\bf r}}-{\rm {\bf 
r}}_{1} } \right|}+\frac{1}{N-1}\left( {v_{ext} ({\rm {\bf r}})+v_{ext} 
({\rm {\bf r}}_{1} )} \right)} \right\}, 
\end{eqnarray}
where the spin orbital $\chi _{\delta } (x)$ is 
the solution of Eq. (\ref{eq21}), 
and should be determined in a self-consistent way. Here note that the 
right-hand side of Eq. (\ref{eq21}) comes from the second term of 
Eq. (\ref{eq20}), and explicitly depends on the spin orbital 
$\chi _{\delta } (x)$. The key point to get the self-consistent solution 
is that spin orbitals of the previous 
iteration are used in calculating $F({\rm {\bf r}})$ and $G_{\delta } 
(x)$\cite{32}. By solving simultaneously Eqs. (\ref{eq19}) and (\ref{eq21}) 
with this technique, 
we can get a new set of spin orbitals and energy parameters $\tilde 
{\varepsilon }_{\delta } \mbox{'s}$. We continue such a procedure until the 
self-consistency for the solutions is accomplished\cite{32}.

\section{CONCLUDING REMARKS}
In this paper, we propose the PD functional theory that yields the best PD 
within the set of PDs that are constructed from the correlated wave 
functions. Compared to the previous one\cite{10,11}, 
the present theory has the following features.

\begin{enumerate}
\item The present theory is superior to the previous one in that 
the searching region of the PDs is certainly larger than the set of 
SSD-representable PDs without missing the merits of the previous 
theory\cite{10}. This means that the resultant PD is more reasonable 
than that of the previous theory. 
\item The predominance of the present scheme can also be shown from 
the viewpoint of the total energy. If the correlation function is chosen 
to be unit, then the present theory is reduced to the previous one exactly. 
It has been already proved that the total energy of the previous theory is 
better than that of the Hartree-Fock approximation\cite{11}. If the 
correlation function is chosen most appropriately, then the searching 
region is substantially equivalent to the set of PDs which are calculated 
by varying both correlation function and spin orbitals in Eq. (\ref{eq20}). 
Therefore, the total energy of the present scheme is necessarily more sound 
than the previous one\cite{10}, and needless to say, 
than that of the Hartree-Fock approximation.
\item In addition to the above merits, the present scheme has the feature 
that deserves special emphasis. Due to the fact that the PD functional 
theory is still a developing field, there hardly exist the computational 
approaches so far. Our previous work is perhaps the first paper to propose 
a computational approach that incorporates both of problems related to 
PD functional theory\cite{10,11}. The present scheme is also a computational 
approach. The resultant simultaneous equations are quite tractable, 
as well as the previous one\cite{10,11}.  Also from such a viewpoint, 
the present scheme seems to be valuable.
\end{enumerate}

Thus, the resultant simultaneous equations (\ref{eq18}) and 
(\ref{eq19}) yield the PD which is definitely closer to 
the ground-state PD than the previous theory\cite{10}. 
Next step is to perform the actual calculations so as to confirm to what 
extent the present scheme covers the $N$-representable PDs. 

Finally, we would like to comment on the future prospect of the present 
theory. Although the present scheme utilizes the lowest-order approximation 
of the expectation value of the PD operator, the higher-order corrections 
can proceed systematically 
with the aid of the Yvon-Mayer diagrams\cite{28,29}. 
Of course, it is anticipated that the equations will become more 
complicated. But, from the methodological point of view, it is important 
that the theoretical framework has the potentiality to improve the 
approximation systematically.

\begin{acknowledgments}
This work was partially supported by a Grant-in-Aid for Scientific Research 
in Priority Areas "Development of New Quantum Simulators and Quantum Design" 
of The Ministry of Education, Culture, Sports, Science, and Technology, 
Japan.
\end{acknowledgments}


\begin{thebibliography}{00}
\bibitem{1}
P. Ziesche, Phys. Lett. A \textbf{195}, 213 (1994).

\bibitem{2}
P. Ziesche, Int. J. Quantum Chem. \textbf{60}, 1361 (1996).

\bibitem{3} 
A. Gonis, T. C. Schulthess, J. van Ek and P. E. A. Turchi, Phys. Rev. 
Lett. \textbf{77}, 2981 (1996).

\bibitem{4}
A. Gonis, T. C. Schulthess, P. E. A. Turchi and J. van Ek, Phys. Rev. B 
\textbf{56}, 9335 (1997).

\bibitem{5} 
A. Nagy, Phys. Rev. A \textbf{66}, 022505 (2002). 

\bibitem{6} 
A. Nagy and C. Amovilli, J. Chem. Phys. \textbf{121}, 6640 (2004).

\bibitem{7}
P. W. Ayers, J. Math. Phys. \textbf{46}, 062107 (2005).

\bibitem{8}
P. W. Ayers and M. Levy, J. Chem. Sci. \textbf{117}, 507 (2005).

\bibitem{9}
P. W. Ayers, S. Golden and M. Levy, J. Chem. Phys. \textbf{124}, 054101 
(2006).

\bibitem{10}
M. Higuchi and K. Higuchi, Physica B \textbf{387} (2007), in press.

\bibitem{11}
M. Higuchi and K. Higuchi, J. Magn. Magn. Mater., in press.

\bibitem{12}
M. Higuchi and K. Higuchi, Phys. Rev. B \textbf{69}, 035113 (2004).

\bibitem{13}
K. Higuchi and M. Higuchi, Phys. Rev. B \textbf{69}, 165118 (2004).

\bibitem{14}
K. Higuchi and M. Higuchi, Phys. Rev. B \textbf{71}, 035116 (2005).

\bibitem{15}
R. G. Parr and W. Yang, \textit{Density-Functional Theory of Atoms and Molecules} (Oxford University Press, New York, 1989) 
Chap. 2.

\bibitem{16}
\textit{Many-electron densities and reduced density matrices}, edited by J. Closlowski (Kluwer Academic/Plenum Publishers, New York, 
2000).

\bibitem{17}
A. J. Coleman, \textit{The Force Concept in Chemistry}, edited by B. M. Deb (Van Nostrand Reinhold, New York, 
1981) p.418. 

\bibitem{18}
A. J. Coleman, Rev. Mod. Phys. \textbf{35}, 668 (1963).

\bibitem{19}
A. J. Coleman and V. I. Yukalov, \textit{Reduced Density Matrices: Coulson's Challenge }(Springer-Verlag, Berlin, 2000). 

\bibitem{20}
\textit{The fundamentals of electron density, density matrix and density functional theory in atoms, molecules and the solid states}, edited by N. I. Gidopoulos and S. Wilson (Kluwer Academic Press, New 
York, 2003).

\bibitem{21}
E. R. Davidson, Chem. Phys. Lett. \textbf{246}, 209 (1995).

\bibitem{22}
S. Kh. Samvelyan, Int. J. Quantum Chem.\textbf{65}, 127 (1997).

\bibitem{23}
M.-E. Pistol, Chem. Phys. Lett. \textbf{400}, 548 (2004). 

\bibitem{24}
M.-E. Pistol, Chem. Phys. Lett. \textbf{417}, 521 (2006).

\bibitem{25}
P. W. Ayers and E. R. Davidson, to appear in Int. J. Quantum Chem. 
(2006).

\bibitem{26}
P. W. Ayers and M. Levy, Chem. Phys. Lett.\textbf{ 415}, 211 (2005).

\bibitem{27}
Since the necessary and sufficient conditions for the $N$-representable 
PDs have not yet been known, we can not judge whether the set of the 
SSD-representable PDs is too small or not. Thus, we may expect that the 
searching the best PD within the set of SSD-representable PDs would give a 
good approximation, while we also have the negative possibility such that 
the searching region would be extremely smaller than the set of 
$N$-representable PDs. In order to remove such a negative possibility, we 
attempt to search the best solution within the PDs that are constructed from 
the correlated wave functions in this paper.

\bibitem{28}
M. D. Ri, S. Stringari and O. Bohigas, Nucl. Phys. \textbf{A376}, 81 
(1982).

\bibitem{29}
M. Gaudin, J. Gillespie and G. Ripka, Nucl. Phys. \textbf{A176}, 237 
(1971).

\bibitem{30}
Ref. 8 and references therein.

\bibitem{31}
M. Levy and P. Ziesche, J. Chem. Phys.\textbf{ 115}, 9110 (2001).

\bibitem{32}
For instance, see, S. E. Koonin, \textit{Computational Physics }(Addison-Wesley, NY, 1986) Chap. 3.
\end{thebibliography}

\end{document}